# Rotationally Warm Molecular Hydrogen in the Orion Bar

Gargi Shaw[1], G. J. Ferland[2], W. J. Henney[3], P. C. Stancil[4], N. P. Abel[5], E.W. Pellegrini[6], J.A. Baldwin[7], & P. A. M. van Hoof[2],[8]

**Abstract**

The Orion Bar is one of the nearest and best-studied photodissociation or photon-dominated regions (PDRs). Observations reveal the presence of $H_2$ lines from vibrationally or rotationally excited upper levels that suggest warm gas temperatures (400 to 700 K). However, standard models of PDRs are unable to reproduce such warm rotational temperatures. In this paper we attempt to explain these observations with new comprehensive models which extend from the $H^+$ region through the Bar and include the magnetic field in the equation of state. We adopt the model parameters from our previous paper which successfully reproduced a wide variety of spectral observations across the Bar. In this model the local cosmic-ray density is enhanced above the galactic background, as is the magnetic field, and which increases the cosmic-ray heating elevating the temperature in the molecular region. The pressure is further enhanced above the gas pressure in the $H^+$ region by the momentum transferred from the absorbed starlight. Here we investigate whether the observed $H_2$ lines can be reproduced with standard assumptions concerning the grain photoelectric emission. We also explore the effects due to the inclusion of recently computed $H_2$ + $H_2$, $H_2$ + H and $H_2$ + He collisional rate coefficients.

*Subject headings:* ISM : molecules — ISM: individual (Orion Bar)

[1] Department of Astronomy and Astrophysics, Tata Institute of Fundamental Research, Mumbai-400-005, India; gargishaw@gmail.com
[2] Department of Physics and Astronomy, University of Kentucky , Lexington, KY 40506; gary@pa.uky.edu
[3] Centro de Radioastronomía y Astrofísica, UNAM Campus Morelia, Apartado Postal 3-72, 58090 Morelia, Michoacán, Mexico; w.henney@astrosmo.unam.mx
[4] Department of Physics and Astronomy and Center for Simulational Physics, University of Georgia, Athens, GA 30602; stancil@physast.uga.edu
[5] College of Applied Science, Department of Mathematics & Physics, University of Cincinnati, Cincinnati, OH, 452061; npabel2@gmail.com
[6] Physics and Astronomy Department, Michigan State University, 3270 Biomedical Physical Sciences Building, East Lansing, MI; pelleg10@pa.msu.edu
[7] Physics and Astronomy Department, Michigan State University, 3270 Biomedical Physical Sciences Building, East Lansing, MI; baldwin@pa.msu.edu
[8] Royal Observatory of Belgium, Ringlaan 3, 1180 Brussels, Belgium; p.vanhoof@oma.be

# 1. Introduction

Molecular hydrogen (H$_2$) is the most abundant molecule in the universe and it accounts for much of the gas in Galactic and extragalactic interstellar clouds. Much of the of H$_2$ emission originates in photodisssociation regions (PDRs), which define a transition zone between an ionized H$^+$ region and cool molecular gas. Electronic fluorescence or photon-pumping causes the temperature derived from line ratios involving high H$_2$ levels to be higher than the kinetic temperature, whereas the temperature derived from line ratios involving only low H$_2$ levels traces the kinetic temperature (van Dishoeck & Black 1986). However, in many PDRs, the temperature derived from the low-level H$_2$ line ratios is also higher than that derived from atomic fine structure line ratios or the kinetic temperature (Wolfire et al. 2003, Timmermann et al. 1996, Fuente et al. 1999, Habart et al. 2003; 2004). Allers et al. (2005; hereafter A05) argue that the high H$_2$ temperature in the Orion Bar measures the kinetic temperature and note that it is difficult to account for the high derived temperature.

Recently Pellegrini et al. (2008, hereafter P08) extensively studied the Orion-Bar using multi-wavelength (X-ray, optical, and IR) data. Their final model reproduced a wide variety of spectral observations across the Orion-Bar and argued that the Bar has significant magnetic pressure together with an enhanced cosmic-ray ionization rate.

The presence of a strong magnetic field is very common in star-forming regions. Earlier Houde et al. (2004) have reported the presence of a magnetic field in Orion A and Rao et al. (1998) studied the magnetic field structure of Orion using linearly polarized dust emission and reported a magnetic field ≤ 35 μG in the H$^+$ region. In addition to this, Abel et al. (2006) also discussed the B field in the Veil, two layers of atomic gas in front of the nebula. So the inclusion of a magnetic field in the P08 model is well justified.

This paper is a sequel to the paper by P08. Here we calculate the rotational temperature of H$_2$ for the Bar, within the context of the P08 model. The challenge is to explain the rotationally warm H$_2$ observed in PDRs.

Several investigators have had difficulty in accounting for the observed temperatures, and have suggested different sources of extra heating. Parmar et al. (1991) reported an excitation temperature of 400-1000 K from emission of 0-0 S(1) and 0-0 S(2) lines in the Orion-Bar. A05 recently derived various H$_2$ excitation temperatures across the Orion-Bar (see their Table 4) and, assuming these to be in local thermodynamical equilibrium with the surrounding gas, showed that standard grain photoelectric emission physics could not produce enough heating to account for the temperature. They used the I[0-0 S(1)]/I[0-0 S(2)] and I[0-0 S(4)]/I[0-0 S(2)] line ratios to calculate excitation temperatures of 400-700 K. The lines are due to optically thin spontaneous radiative transitions and so the line intensities are proportional to the column density of the upper level. Hence, the line ratio I[0-0 S(1)]/I[0-0 S(2)] is proportional to $N(0,3)$ and $N(0,4)$ where $N(v,J)$ corresponds to the column density of the $v, J$ rovibrational state. The excitation temperature $t_{ul}$ is defined from the ratios of column densities $N(v,J)$ as

$$t_{ul} = -\frac{\Delta E_{ul}/k}{\ln\left[\dfrac{N(v_u,J_u)/g_u}{N(v_l,J_l)/g_l}\right]} \quad [\text{K}]. \tag{1}$$

where $u$ and $l$ are the upper and lower states, respectively. So the excitation temperature $t_{43}$ is derived from the ratios of the column densities of $N(0,4)$ and $N(0,3)$. Similarly, the $t_{64}$ excitation temperature derived from I[0-0 S(4)]/I[0-0 S(2)] is proportional to the column density ratios of $N(0,6)$ and $N(0,4)$. If photon pumping is important then the level populations will be super-thermal as discussed below. If these levels are thermalized then $t_{ul}$ is equal to the gas kinetic temperature and we find the conundrum posed by A05.

Recently Henney et al. (2007) have studied the merged ionization / dissociation front in the Helix nebula and reproduced the high observed luminosity in $H_2$ lines together with the high excitation temperature with the help of steady state dynamics. We do not consider dynamics in the current models, which is justified because the direct effects of dynamics on the global emission properties are likely to be much smaller in the Orion Bar than in the Helix knots, due to the much higher ionization parameter (Henney et al 2005a).

In this paper we show the spatial variation of several $H_2$ level populations across the Orion-Bar and predict $t_{43}$ and $t_{64}$ together with the intensities of 1-0 S(1), 0-0 S(2) and 0-0 S(4) lines. We perform numerical simulations with the recently developed model of $H_2$ that is part of the spectral simulation code Cloudy (Shaw et al. 2005) and explicitly include the $H^+$ region as part of the calculation. This allows us to include the starlight momentum in the gas equation of state.

## 2. Calculations

All numerical calculations are performed using the version 08.00 of the numerical spectral simulation code Cloudy, last described by Ferland et al. (1998). This code is based on a self-consistent calculation of the thermal, ionization, and chemical balance of gas and dust exposed to a source of radiation. Our chemical network consists of ~$10^3$ reactions with 71 species containing hydrogen, helium, carbon, nitrogen, oxygen, silicon, sulfur, and chlorine (Abel et al. 2005). Further, we include silicate and graphite size-resolved grains and determine the grain charges and photoelectric heating self-consistently (van Hoof et al. 2004).

## *2.1. The hydrogen molecule*

Our treatment of $H_2$ is described in Shaw et al. (2005, hereafter S05; 2008) and is briefly discussed here. We use a detailed $H_2$ chemistry network, consisting of various state-specific formation and destruction processes. We include all 301 bound levels within the ground electronic state, and all levels within the lowest six electronic excited states which are coupled to the ground electronic state by permitted electronic transitions. $H_2$ is predominantly formed on dust grains in the ISM. Theoretical (Cazaux &Tielens 2002) and observational (Habart et al. 2004) studies of $H_2$ formation on grain surfaces are available. As these results are consistent, we adopted the $H_2$ formation model of Cazaux & Tielens (2002). We also consider $H_2$ formation via associative detachment ($H^- + H^0 \rightarrow H_2 + e^-$). These exothermic formation processes produce $H_2$ in excited vibrational and rotational levels, often referred to as formation pumping. $H_2$ is destroyed mainly via the Solomon process (electronic fluorescence) when the gas is optically thin to the $H_2$ electronic lines. Most electronic excitations of $H_2$ result in decays to highly excited vibrational and rotational levels of the ground electronic state (Solomon pumping), which further decay via quadrupole transitions (Tielens & Hollenbach 1985, Black & van Dishoeck 1987, Abgrall et al. 1992). $H_2$ can also be collisionally dissociated from high vibrational levels. We also consider excitation to the triplet and singlet (B, C, B´, and D) states via secondary electrons produced by cosmic rays following Dalgarno et al. (1999) and Liu & Dalgarno (1994); this effect eventually populates the vibrational and rotational levels of the ground electronic state. For further details see section 2.1 and eqn. 2 of Shaw et al. (2008). Radiative decays between ortho- and para- $H_2$ states are not possible because of their different nuclear spin. However, exchange collisions with H, $H^+$, and $H_3^+$ can convert ortho-$H_2$ into para-$H_2$. As a result of these processes (for which we have included all available data), $H_2$ will not be in local thermodynamic equilibrium (LTE) with the surrounding gas in the Solomon-dominated region where its electronic lines are optically thin. Therefore, the ortho-para ratio (and the purely rotational $H_2$ line emission) may not trace the kinetic temperature in many astrophysical environments.

Collisions play an important role in determining the level populations and so it is always vital to use accurate collisional rate coefficients. We have incorporated new data for several collisional processes since the publication of S05. In this paper we use the updated non-reactive $H_2 + H$ collisional rates from Wrathmall et al. (2007) which are orders of magnitude larger than the previously used data from Le Bourlot et al. (1999) for vibrational transitions. Though the calculations of Wrathmall et al. (2007) are quantum mechanical and adopted the most recent $H_3$ potential energy surface, uncertainty still remains due to their neglect of reactive scattering processes. In addition, we also use updated $H_2 + He$ collisional data from Lee et al. (2005, 2008b). Recently Lee et al. (2006, 2008a) have performed an extensive quantum mechanical calculation for $H_2 + H_2$ collisions using a reliable potential energy surface (Diep & Johnson 2000). Their rate-coefficients show significant differences with those adopted in Le Bourlot et al. (1999), particularly below 200 K. In Cloudy, an option has been added to allow for the selection of the new Lee et al. (2008a) rate coefficients.

## 2.2. Cloud geometry

Classically, a PDR is defined as a region where hydrogen is in atomic or molecular form and is illuminated by far ultraviolet (FUV) (6 eV < hν < 13.6 eV) photons. However, these regions are not isolated; they are adjacent to an H II or $H^+$ region if the central star is hot enough. Therefore, the ionized region, PDR, and molecular cloud are coupled to each other via proximity and dynamics. So, instead of the classical-isolated PDR model, we simulate the full environment as it occurs in nature – the incident continuum propagates from the $H^+$ region into the molecular cloud through the PDR. The radiative transport of the full continuum, from radio through x-rays, is followed including line overlap for the hundreds of thousands of electronic transitions that determine the excitations and destruction of $H_2$. Further details are given in Abel et al. (2005), while Kaufmann et al. (2006) have performed a similar type of calculation.

The Orion-Bar PDR is located in the Orion molecular cloud ~2′ southeast of the Trapezium cluster. Figure 1a shows a cross-section through the Orion-Bar perpendicular to the plane of the sky. The Bar is seen nearly edge on and so has been an important test of PDR physics (Osterbrock & Ferland 2006; section 8.5).

We calculate one-dimensional, static models in a plane-parallel slab geometry. We assume that the FUV radiation emitted by the Trapezium Cluster is dominated by $\theta^1$ Ori C. These models are an approximation to the physical structure of the Bar along an extension of the dotted line in Figure 1a marked "0.114 pc".

To model the geometry shown in Figure 1a we calculate the physical properties along the direction of the radiation field and then integrate this along the line of sight through the cloud. This 1D approximation is an obvious oversimplification but one that is common in simulations of irradiated molecular gas. As shown in Figure 1a, the simplest model of the Bar is a slab viewed from a particular angle (Wen & O'Dell 1992; Henney et al. 2005b, Walmsley et al. 2000). Figure 3 of P08 shows the idealized geometry used in our calculation. Once the model is calculated the predicted surface brightnesses of the emission lines can be calculated by integrating over the geometry. We take the geometry to be a tilted slab with line of sight length $h$, angle $\theta$, and a radial distance $r$ from the ionizing source $\theta^1$ Ori C. Then, using the volume emissivity of each line predicted by Cloudy as a function of $r$ from $\theta^1$ Ori C, we step along the $x$ axis in intervals of $dx$ and integrated the emissivity $\varepsilon(r(x,y))$ along the line of sight depth $y$ in steps of $dy$. The contribution of the emissivity to the observed surface is corrected for internal reddening through the cloud along $y$ during the integration using a dust reddening law consistent with the Orion Nebula.

## *2.3. Model parameters*

The geometry within the inner regions of the Orion Nebula is well constrained by observations. The equation of state, the relationship between the gas density, temperature, and depth into the Bar, is critical since the calculations will extend from hot ($T \approx 10^4$ K) ionized regions, into the warm predominantly atomic PDR, and eventually into cold molecular gas. Our adopted model parameters are taken from P08 and are listed in Table 1. We take the number of ionizing photons emitted by the star to be $Q(H) = 10^{48.99}$ s$^{-1}$ as described in P08. The stellar continuum is based on a Kurucz model with $T_{eff}$ = 39,700 K modified to reproduce the [Ne III] and near IR lines observed in the Orion nebula (Rubin 1991). We include the effects of X-rays by considering bremsstrahlung at $10^6$ K with an integrated luminosity $L_x = 10^{32.6}$ erg s$^{-1}$ over the 0.5–8 keV passband (Feigelson et al. 2005). The electron density within the illuminated H$^+$ layer of the Bar is measured to be $n_e \approx 10^{3.2}$ cm$^{-3}$ (Pogge et al. 1992). Our grain type is the same as given in Baldwin et al. (1991) with $R_v$ = 5.5. We use a modified MRN (Mathis et al. 1977) size distribution of grains, i.e, $dn/da \propto a^{-3.5}$ where $a$ is the radius of the grain and $n$ is the number of grains with a radius between $a$ and $a+da$. In our model we consider 10 size bins with $a_{min}$ = 0.03 μm and $a_{max}$ = 0.25 μm. The mean size of these grains is larger than the standard MRN grains with $R_v$ = 3.1 due to the lack of small particles. We also include PAHs ($n_{PAH}/n_H^0 = 3\times10^{-7}$) with a power-law distribution of PAH sizes with 10 size bins, according to Abel et al. (2008).

The magnetic field strength in Orion's Veil is known to be large (Abel et al. 2004) and this can affect the physical conditions. Based on that we include a magnetic field $B = B_0 \times \left(\frac{n}{n_0}\right)^{\gamma/2}$ where $B_0$ and $n_0$ are the magnetic field and the gas density at the illuminated face of the cloud and γ depends on the geometry of the system. We use γ =2 in our model which corresponds to plane parallel compression perpendicular to the field lines. Our model includes enhanced cosmic ray effects assuming that cosmic rays are trapped in the magnetic field and their density is enhanced along with the magnetic field strength. The cosmic-ray density is given by $n(cr, B) = n(cr, B_0)\left(\frac{U(B)}{U(B_0)}\right)$ cm$^{-3}$, where $U(B)$ and $U(B_0)$ are the energy densities of the local magnetic field relative to the galactic background magnetic field. The heating and ionization effects of the cosmic-rays are included accounting for this energy density and the local ionization fraction of the gas. We also include the cosmic microwave background (CMB).

We use the gas-phase abundances listed in Table 2, which are primarily derived from observations of the Orion Nebula by Baldwin et al. (1991), Rubin et al. (1991, 1993) and Osterbrock et al. (1992). The Ne abundance found by Baldwin et al. (1991) was affected by the stellar atmosphere used in that paper, so we have adopted the Osterbrock et al.(1992) value.

### *2.4. A ray into the Bar*

Figures 1b and 1c show the hydrogen-ionization structure without and with an enhanced cosmic-ray ionization rate, respectively. The effect of the enhancement is most prominent at the shielded face of the cloud and as a natural consequence of increased ionization of $H_2$, the density of $H^+$ and $H^0$ are larger at the shielded face. Figures 1d and 1e show the temperature profile, and various heating mechanisms without and with an enhanced cosmic-ray ionization rate, respectively. At the shielded face of the enhanced cosmic-ray ionization model, heating is dominated by cosmic ray ionization and the temperature is also higher. Similarly, figures 2a and 2b show the variation of the kinetic temperature, the $H_2$ density, and its various levels across the cloud with and without an enhanced cosmic-ray ionization rate, respectively. For a model in the absence of enhanced cosmic-ray ionization rate, $H_2$ is less ionized and as a result the molecular fraction is larger and more sharply peaks with depth (compare figures 1b and 1c) giving rise to a narrower $H_2$ spectral line width. As cosmic rays penetrate deeper, an enhanced cosmic-ray ionization rate will keep the molecular fraction low for a larger distance giving rise to a broader $H_2$ line width. However, the actual geometry of the Bar is more complex than we have assumed so subtle differences in the width may be due to details of the geometry which are beyond the current model. The gas temperature profile in Allers et al. (2005) (their fig 6.) and the current work (fig 2a) are similar. We also show the predicted variation of the total ortho-to-para ratio across the cloud in Figure 2c. At shallower depth this ratio is ≈ 3 and as self-shielding starts to build up it increases to a value ~ 7 due to the preferential self-shielding of ortho-states (Sternberg & Neufeld 1999) and then goes below the value 3 in deeper regions. The surface brightness distributions in other important lines like Hα are shown in fig 5. of P08. The cosmic ray rate in this model is not a constant value; it increases along with the magnetic field since we assume that the two are in equipartition. At the position $5.05 \times 10^{17}$ cm (see figure 2a, the $H_2$ lines peak at this depth) the cosmic ray ionization rate is $7 \times 10^{-14}$ $s^{-1}$. This value is ~300 times larger than the background value of $2 \times 10^{-16}$ $s^{-1}$ found by Indriolo et al. (2007). Van der Tak et al. (2006) have also observed an enhanced cosmic ray ionization rate (~$4 \times 10^{-16}$ $s^{-1}$) in the Sagittarious B region.

The kinetic temperature decreases into the cloud and the $H_2$ excitation temperatures also vary across the cloud as the $H_2$ level populations change in response to changes in the photon pumping and collisional rates. The observed intensities of the 0-0 S(2) and 0-0 S(4) lines are proportional to $n(0,4)$ and $n(0,6)$ and hence the surface brightnesses of these lines reach maxima at the peak values of $n(0,4)$ and $n(0,6)$, respectively. Figures 3 and 4 show the excitation temperatures $t_{43}$ and $t_{64}$, and the kinetic temperature as a function of $n(0,4)$ and $n(0,6)$, respectively. Our goal is to predict the rotational temperature at positions where $n(0,4)$ and $n(0,6)$ peak. At the peak position of $n(0,4)$ both the kinetic temperature and $t_{43}$ are ~265 K. However, at the peak position of $n(0,6)$ the kinetic temperature is ~ 223 K and $t_{64}$ is ~556 K. Our predicted surface brightness (taking into account the internal reddening effect) for the 0-0 S(1), 0-0 S(2), and 0-0 S(4) lines are $5.19 \times 10^{-4}$, $1.14 \times 10^{-4}$, and $1.06 \times 10^{-4}$ erg$s^{-1}$cm$^{-2}$sr$^{-1}$ which also give $t_{43}$ = 242.39 and $t_{64}$ = 573.20 K, respectively.

We next consider detailed level populations to understand the differences in the kinetic and $H_2$ excitation temperatures. For simplicity we consider a single ray from the star through the PDR into the molecular cloud. Figure 5a plots the predicted column densities integrated along this ray. This ray extends into the molecular cloud and has a total $H_2$ column density of $1 \times 10^{22}$ cm$^{-2}$. This is considerably larger than the observed $H_2$ column density $9 \times 10^{20}$ cm$^{-2}$, for a line from the Earth through the Bar. The figure also shows the column densities predicted by assuming that level populations are in LTE at every point along the ray. The LTE populations do not lie along a straight line because of changes in the kinetic temperature along the ray. It is clear that the $J = 0–3$ levels are in LTE. The populations of these low levels are mostly affected by collisions and henceforth have come into equilibrium. The higher levels are mainly populated by non-thermal pumping processes and are overpopulated relative to their LTE values. Hence, $t_{64}$ does not trace the kinetic temperature as the corresponding rotational levels are overpopulated. Figure 5b shows the same quantities as in figure 5a, but without an enhanced cosmic ray ionization rate. It is clearly seen that the presence of an enhanced cosmic ray ionization rate provides extra non-thermal pumping to the higher rotational levels (compare the $J = 4$ level).

A05 found that $t_{43}$ was around 450K and $t_{64}$ was around 550K. Figure 5 shows that in our simulations $t_{43}$ traces the kinetic temperature while $t_{64}$ is strongly affected by fluorescence pumping. The predicted $t_{64}$ is reasonable close to the observed value. However $t_{43}$ is around 250K, significantly below the measured value.

The measured variation in $t_{43}$ (390 to 630 K) is a great deal larger than that in $t_{64}$ (500 to 570 K). At the two positions where both measurements exist $t_{64}$ is 500 and 570 K which we reproduce as a non-thermal fluorescent effect. At the same positions $t_{43}$ is 430 and 460 K. If $t_{43}$ and $t_{64}$ both trace the kinetic temperature, as assumed by A05, they would agree. Our calculations are closer to the lower value.

### *2.5. The effect of improvements in the collisional rates*
Collisions play an important role in determining level populations and the uncertainties in the collisional rate coefficients are substantial. As the reliability of potential energy surfaces and the availability of computational power increases, the accuracy of the collisional calculations improves. Some of the more recently computed rate coefficients were found to be significantly different from previous values. We show the effect of these different rates on $t_{43}$ and $t_{64}$ in Figures 6 and 7. Our calculations include both reactive and non-reactive collisions as summarized in S05. The kinetic temperature near the peak of $n(0,4)$ and $n(0,6)$ are ~250 K. Reactive collisional processes are not significant at these temperatures, since the barrier to reactivity for $H_2$-H collisions is ~3900 K. In S05 we used the rate coefficient fits given by Le Bourlot et al. (1999) for non-reactive collisions which include rovibrational excitation of $H_2$ by collisions with H, He, and $H_2$. Later we updated to the $H_2$ + He collisional data from Lee et al. (2005, 2008b) and started to use them by default in Cloudy. We have also updated the $H_2$ + H and $H_2$ + $H_2$ rate coefficients as mentioned earlier. We consider four different models using different combinations of rate coefficients as indicated in Table 3, all other colliders are as given in S05. Table 4 compares various rate coefficients at 500K for the $H_2(0,8)$ to $H_2(0,6)$ transition.

Figure 6 shows that the peak value of $n(0,4)$ is decreased with the new rates (S05 + new He rates, S05 + new He + new H rates, and current rates) compared to S05 rates and at the same time the value of $t_{43}$ near the peak of $n(0,4)$ also is decreased by ~15K with new rates compared to S05 rates (280 K). Similarly, the peak value of $n(0,6)$ also is decreased with the new rates compared to S05 rates. Here the value of $t_{64}$ near the peak of $n(0,6)$ is 465 K with S05 rates and is increased by ~120 K with S05 + new He rates, and by ~100 K with S05 + new He + new H rates, and current rates.

## *2.6. Comparison to Observed Surface Brightness Profiles*

We now compare our results to the A05 observations, which were taken at a location where the $H_2$ emission is bright. The models described so far use the same input parameters as the enhanced Cosmic Ray model from P08, which are appropriate for the typical inter-clump gas in the Bar. The $H_2$ surface brightness at the A05 position is brighter because the gas density is higher. This density enhancement can be seen at the position (RA offset = -29", Dec offset = -105", $V_\odot$ = + 10 km s$^{-1}$) in the density map presented in Figure 5 of Garcia-Diaz & Henney (2007), which is based on the [S II] 6716/6731 intensity ratio. The [S II] density at that spot is log ($n_e$) = 3.9, which corresponds to $I$([S II] $\lambda 6716$)/$I$([S II] $\lambda$ 6731) = 0.53. We ran an additional model with the hydrogen density at the illuminated face increased by 0.3 dex over the value in the P08 model. This new model reproduces the [S II] intensity ratio observed at the A05 position and also produces higher $H_2$ surface brightnesses that are in much closer agreement with the A05 observations than those produced by the lower density model (see Figure 8a). In this "increased density" model, the peak surface brightnesses of the 0–0 S(1), 0–0 S(2) and 1–0 S(1) lines now are in the correct order (from brightest to faintest) i.e., the 0-0 S(2) line is brighter than the 1-0 S(1) line which is consistent with the observations, and straddle the observed peak surface brightnesses, which we consider to be a satisfactory match. The predicted $t_{64}$ and $t_{43}$ for this model is 424 and 280 K. The width of the observed spectral lines are wider than our predictions; suggesting that the cosmic ray rate might be higher than what we have assumed. However, due to the uncertain details of the geometry, it is hard to constrain the cosmic ray rates by comparing the observed and predicted line widths. Figure 8b shows the computed $H_2$ surface brightness of 0–0 S(1), 0–0 S(2), and 1–0 S(1) lines without an enhanced cosmic ray ionization rate. The lines are much sharper as is discussed already in section 2.4. The non-thermal pumping produced by the cosmic rays populates the higher vib-rotational levels more effectively than the lower levels. Hence the increased cosmic-ray ionization rate is more important for the 1-0 S(1) than the 0-0 S(1) and 0-0 S(2) line data fit.

### *2.7. Heating sources*

The challenge in explaining the kinetic temperature of the $H_2$-emitting gas is discussed by Allers et al. (2005). They found, as we do, that conventional photon interactions with dusty atomic/molecular gas cannot produce enough heating to explain the observations. They suggested that the theory of gas heating by electron photoejection from grains may need to be revised. We encountered similar problems in reproducing the observed high gas kinetic temperatures. We have examined the effects of a higher-than-Galactic cosmic ray density and find that this can account for many of the observations.

The Galactic background cosmic-ray ionization rate is undergoing revisions with the development of new molecular diagnostics. Indriolo et al. (2007) argue that the Galactic background in diffuse clouds is ~1 dex more intense than previously thought and has a significant variance over different sight lines, which is consistent with recent theoretical modeling (Padoan, Paolo & Scalo, John 2005; Snodin, Andrew P. et al. 2006). An increased cosmic-ray density in dense molecular gas is harder to justify theoretically and our final value for the Bar is higher than those measured by Indriolo et al. However, in our previous study of M17 (Pellegrini et al 2007) we found a similarly high cosmic ray flux. This was justified in that case since a supernova remnant, thought to be the source of galactic cosmic rays, is very near the M17 cloud. No such supernova is close to the Orion complex.

We offer this as an alternate explanation of the observations, much in the spirit of Allers et al (2005)'s suggestion that grain heating theory may need to be revisited. Our calculations show that enhanced cosmic rays can account for many observed aspects of the Bar, but, given the theoretical difficulties of this scenario that are outlined in the previous paragraph, it is worth investigating alternative explanations. Cosmic rays both heat and ionize the gas, with the relative proportion being set by the ionization fraction of the gas (Osterbrock & Ferland, 2006). Although it is hard to distinguish between the effects of cosmic rays and X-ray photoionization from the $H_2$ spectrum, Meijerink et al. (2006) identify other tracers, including high-$J$ CO lines and [CI]/$^{13}$CO, HCN/CO, and HCN/HCO$^+$ line ratios, which are sensitive to the power source. Maps of the bar in these lines might be decisive. Observationally it is very hard to distinguish between cosmic ray and x-ray photoionization, a point stressed in Ferland et al. (2009). Young stars in the Orion complex are known to be significant sources of x-rays. So a third possibility is that the needed heating is provided by a distributed population of young stars with their associated x-ray emission.

In contrast to the above is heating by grain electron photoejection, in which case the primary electrons are not capable of ionizing the surrounding gas so the energetic electrons are quickly thermalized. This latter process produces heating without associated ionization. A careful analysis which combined measurements of the gas kinetic temperature with observations of the abundances of molecular ions could discriminate between these possible processes.

## 3. Summary & Conclusions

This paper revisits emission properties of the Orion-Bar. Our conclusions include the following.

- We include the $H^+$, $H^0$, and $H_2$ regions in one self consistent calculation and include the stellar radiation pressure, the magnetic field, and an enhanced cosmic-ray ionization rate in the gas equation of state. The model parameters are taken from P08 who successfully reproduced a wide variety of spectral observations across the Bar. We qualitatively reproduce the observed $H_2$ rotational temperature $t_{64}$. There are some uncertainties regarding the magnetic field strength, molecular data, and geometry which may lead to a lower prediction for $t_{43}$.
- Our predicted $t_{43}$ is close to the kinetic temperature (~250K) but our predicted $t_{64}$ is much greater. We find that in the Orion-Bar rotational levels with $J > 3$ are not in thermodynamical equilibrium with the local temperature. The excitation temperature $t_{64}$ is much higher than the kinetic temperature due to photoexcitation (FUV pumping). This provides an explanation for why $t_{64}$ is high.

- We predict that $t_{43}$ is thermal and that it is lower than observed. This model will require an enhanced heating rate in the atomic region of the PDR to increase both $t_{43}$ and the kinetic temperature to 450 K. A05 increased the standard photoelectric heating rate by a factor ~3 to achieve this extra heating.
- The collisional rates for $H_2$ with H and vibrational transitions for $H_2 + H_2$ are uncertain. More accurate collisional rates involving $J=3$ and 4 might bring the predicted $t_{43}$ closer to its observational value. We note that A05 correctly predicted that the vibrationally excited $H_2$–H collisional rates were, at that time, significantly too small.
- We show that a variant on the P08 model, in which the density was increased to match that of the small higher-density clump measured by Garcia-Diaz & Henney in [S II], produces the surface brightness and relative intensities of the $H_2$ 0–0 S(1), 0–0 S(2) and 1–0 S(1) lines.
- The starlight momentum increases the pressure in the Bar above what would be predicted with a constant gas pressure model with the measured gas pressure in the HII region. Some of this pressure is magnetic. That, together with a postulated enhanced cosmic ray density, can account for most, but not all, of the observed $H_2$ emission properties.

## Acknowledgements

PCS acknowledges NASA Grant NNG05GD81G and NSF Grant AST-0607733. NPA would like to acknowledge support from the National Science Foundation under Grant No. 0094050, 0607497 to The University of Cincinnati. PvH acknowledges support from NSF grant AST0607028. JAB gratefully acknowledges support from NSF grant AST-0305833. EWP was supported by NASA (07-ATFP07-0124 and STScI AR-10932) and by NSF (AST-0305833). GJF is funded by NASA (07-ATFP07-0124) and NSF (AST0607028). We would like to thank the referee for thoughtful suggestions.

# 4. Figures

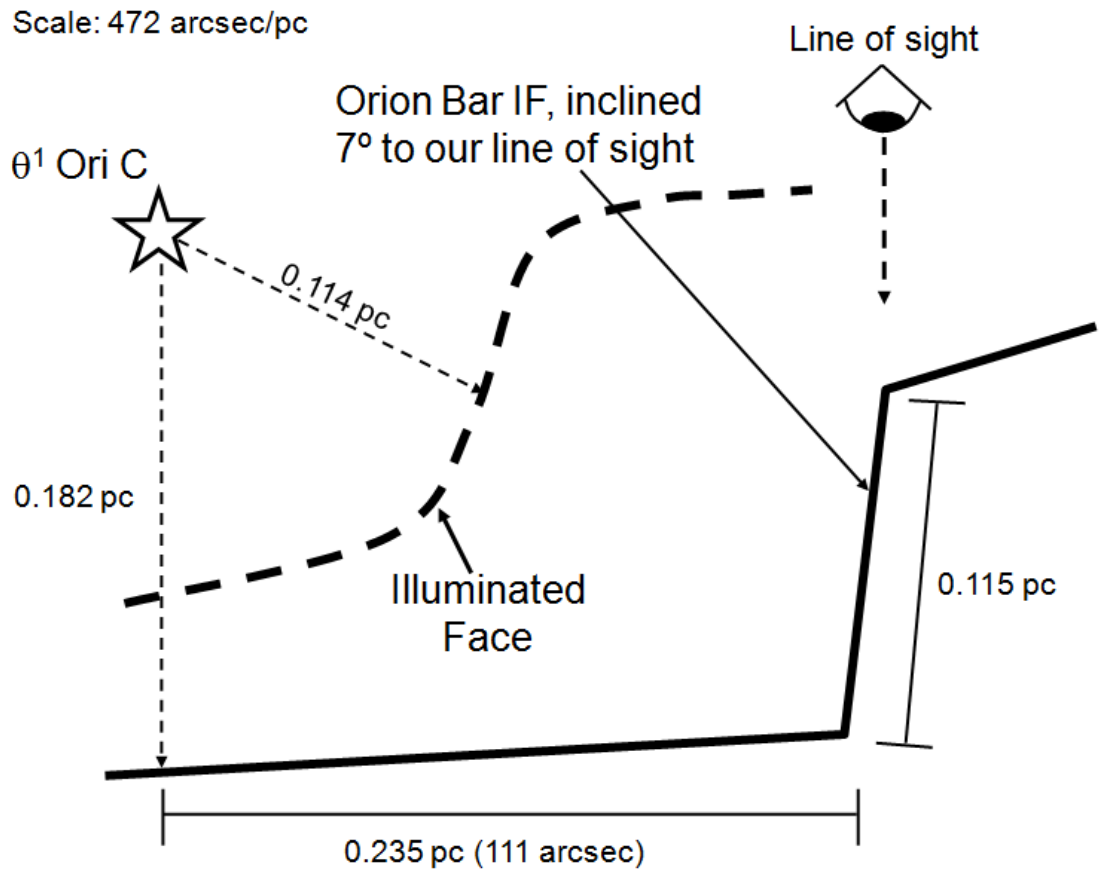

Figure 1a. The simplified geometry adopted here and by P08. The derivation of the distances shown on the figure is described by P08.

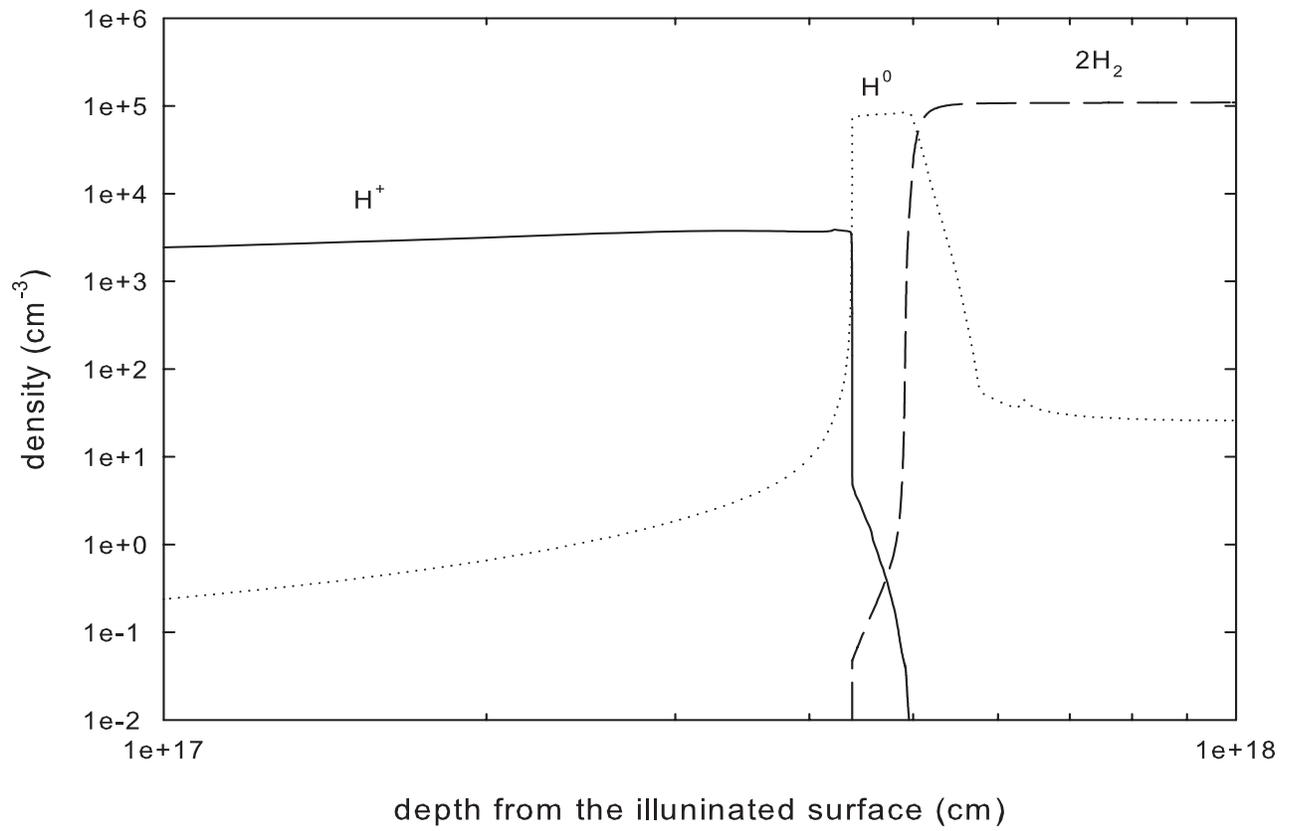

Figure 1b. The hydrogen-ionization structure without an enhanced cosmic-ray ionization rate.

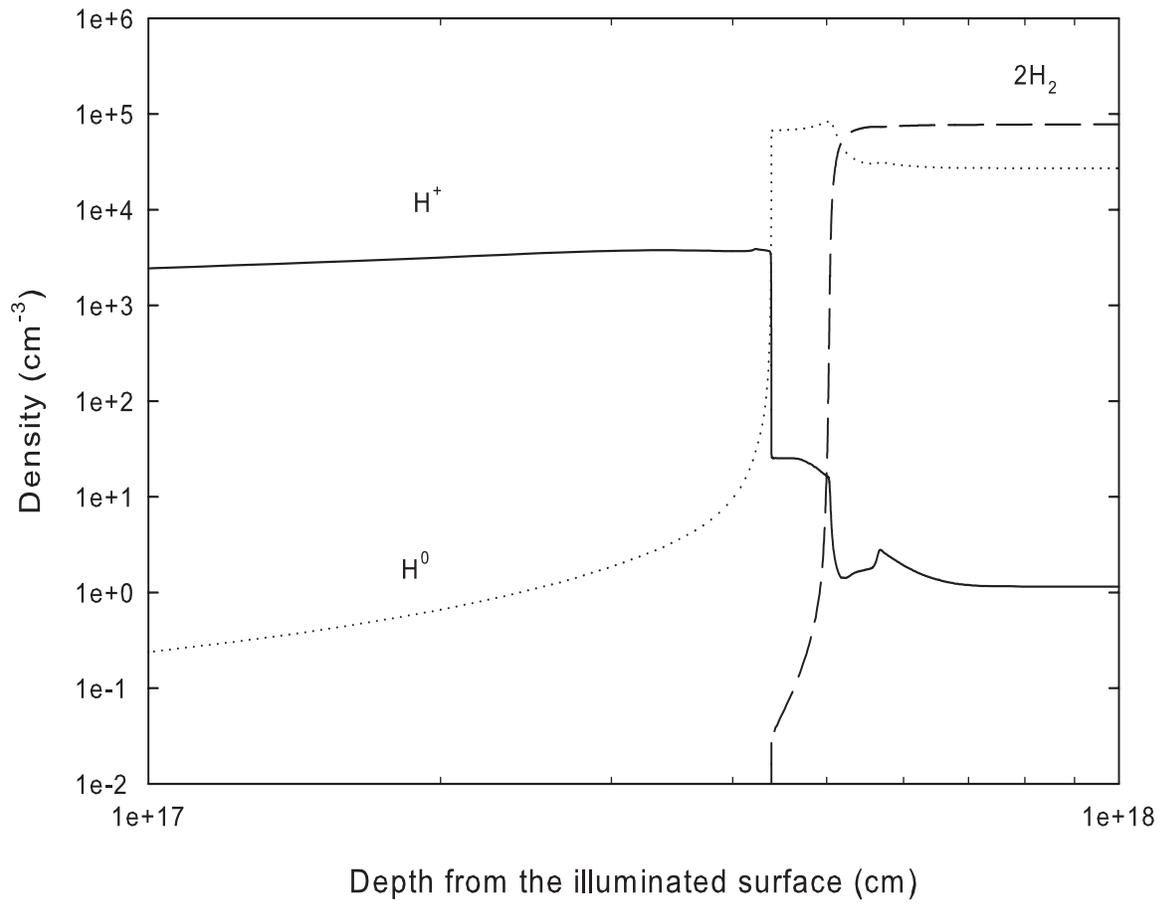

Figure 1c. The hydrogen-ionization structure with an enhanced cosmic-ray ionization rate.

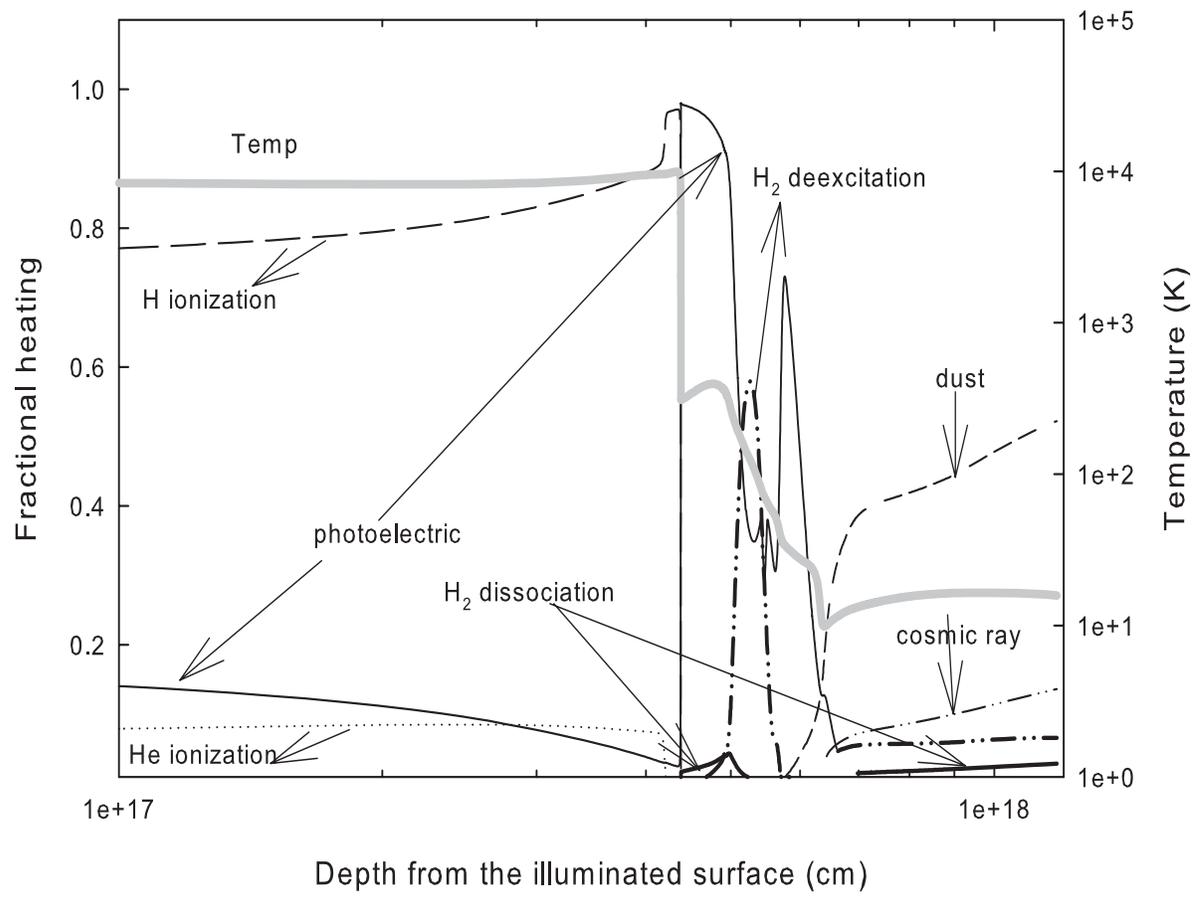

Figure 1d. The temperature profile and various heating mechanisms without an enhanced cosmic-ray ionization rate.

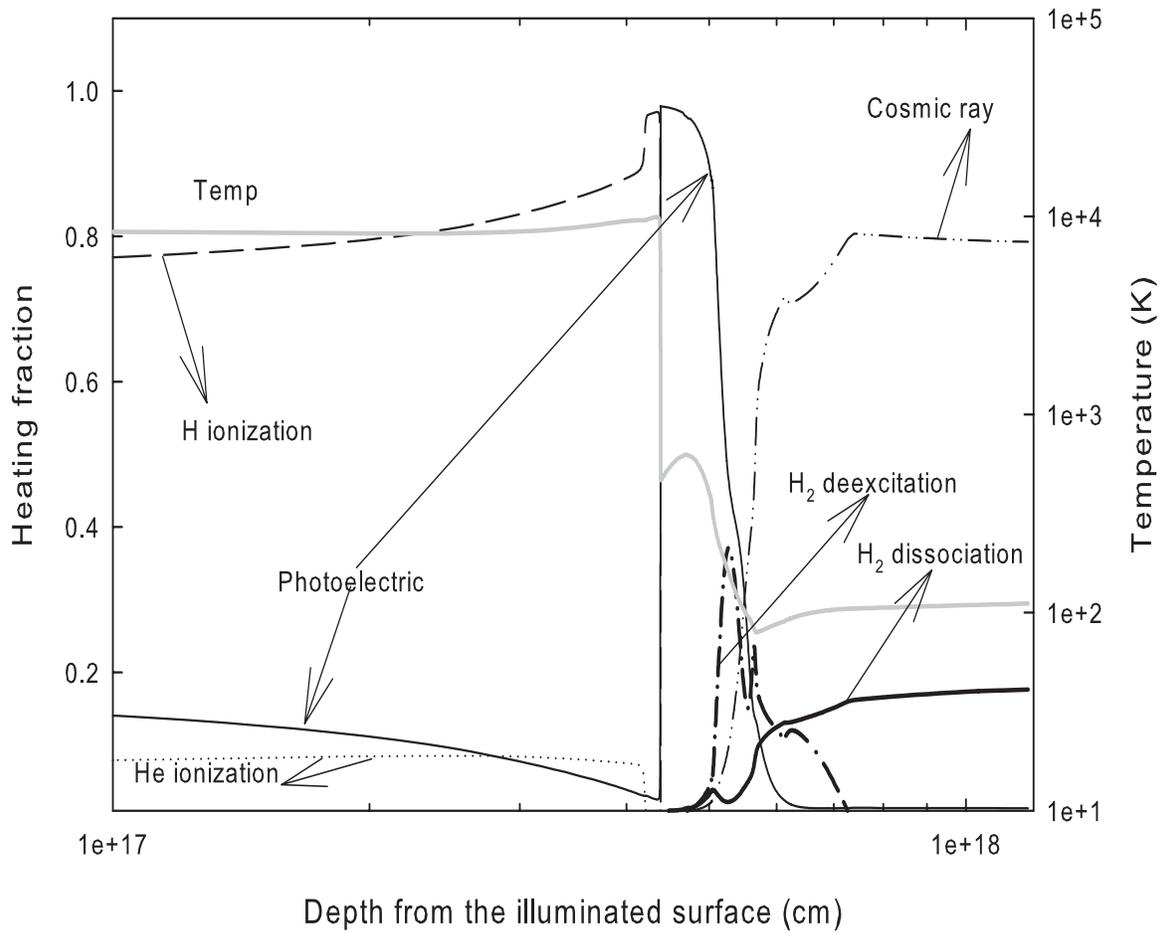

Figure 1e. The temperature profile and various heating mechanisms with an enhanced cosmic- ray ionization rate.

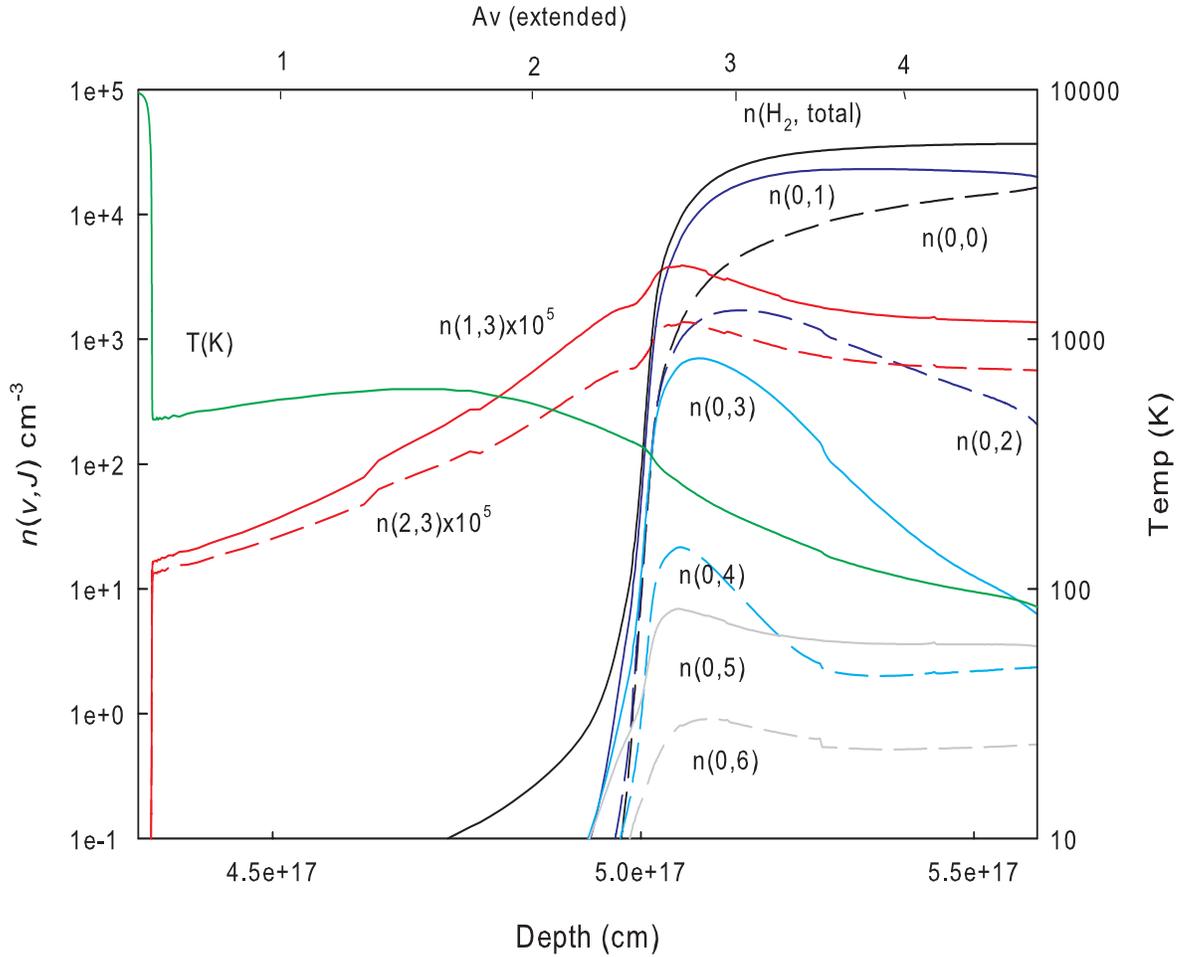

Figure 2a. Variation of the kinetic temperature, the total $H_2$ density, and the density of some $H_2$ rovibrational levels $(v, J)$ across the cloud. The variables jump discontinuously at the ionization front (left end). This and Fig. 2c show the variation along a single ray through the Bar.

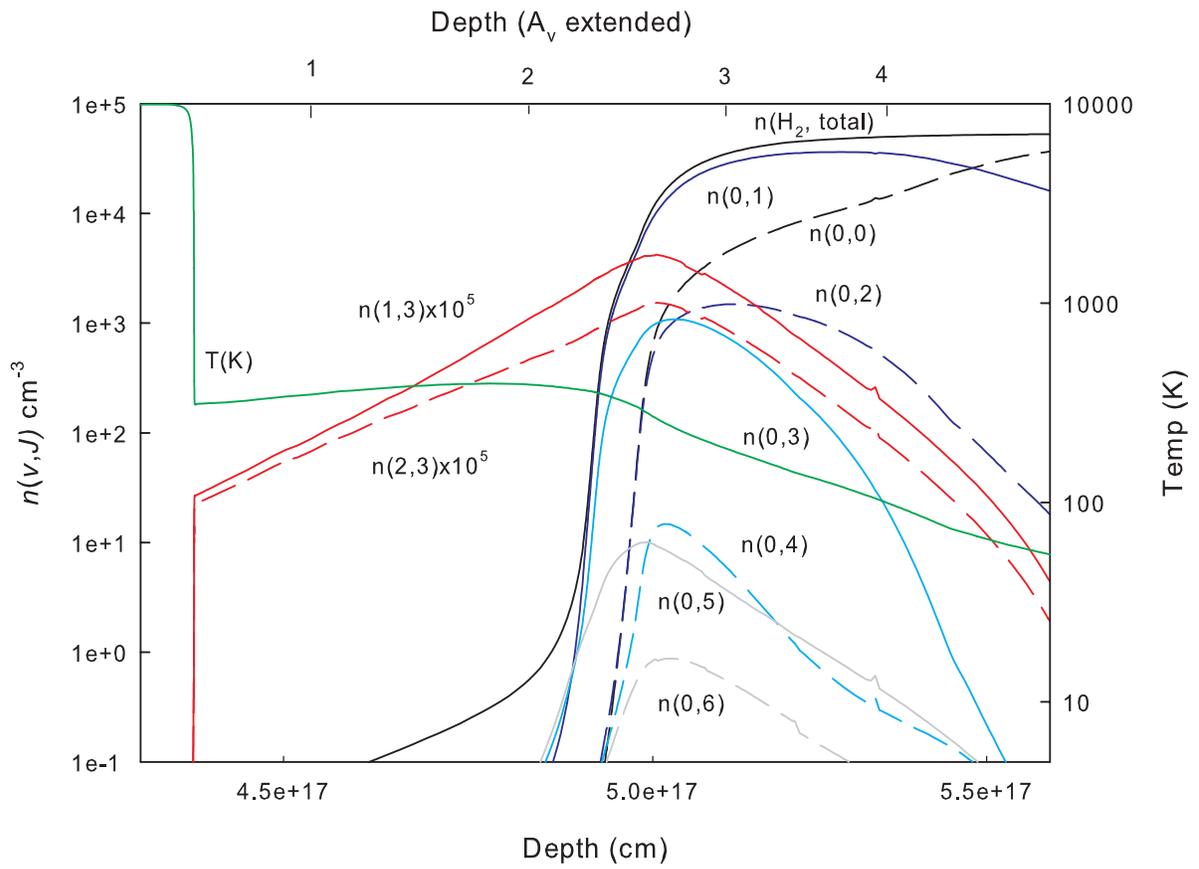

Figure 2b. Same as figure 2a, but without an enhanced cosmic-ray ionization rate.

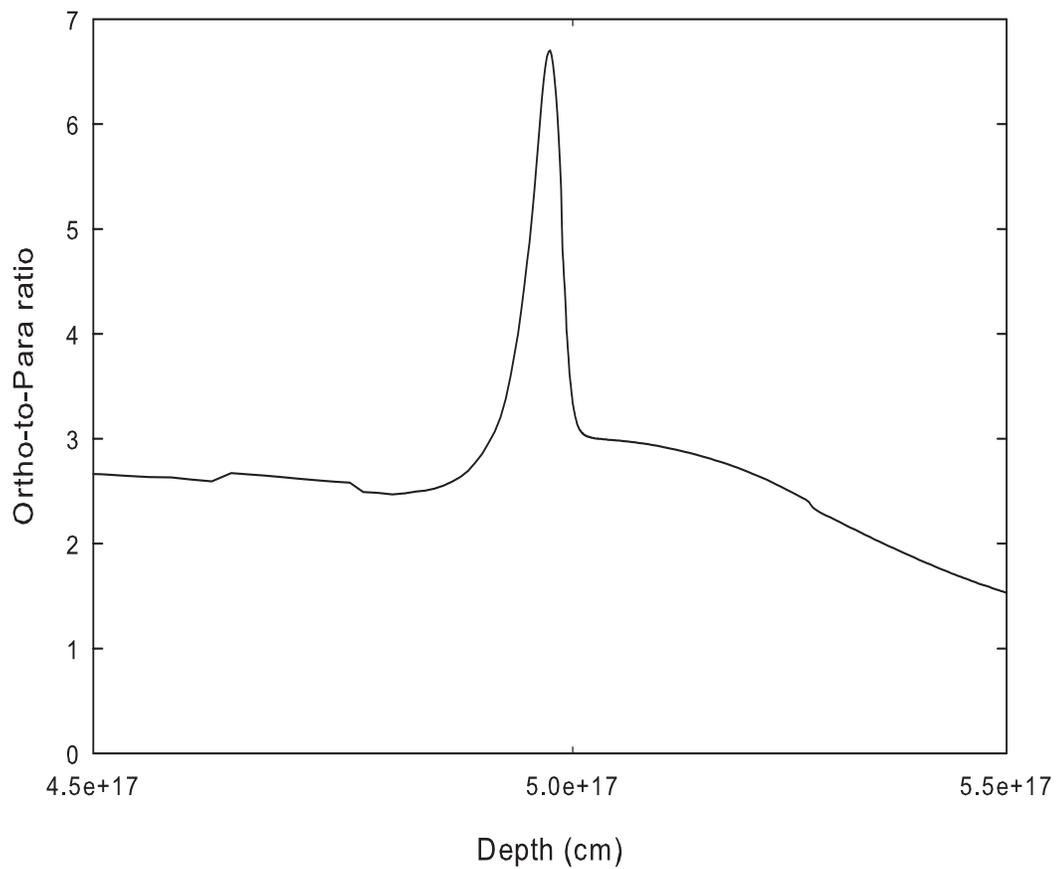

Figure 2c. Variation of the Ortho-to-Para ratio across the cloud.

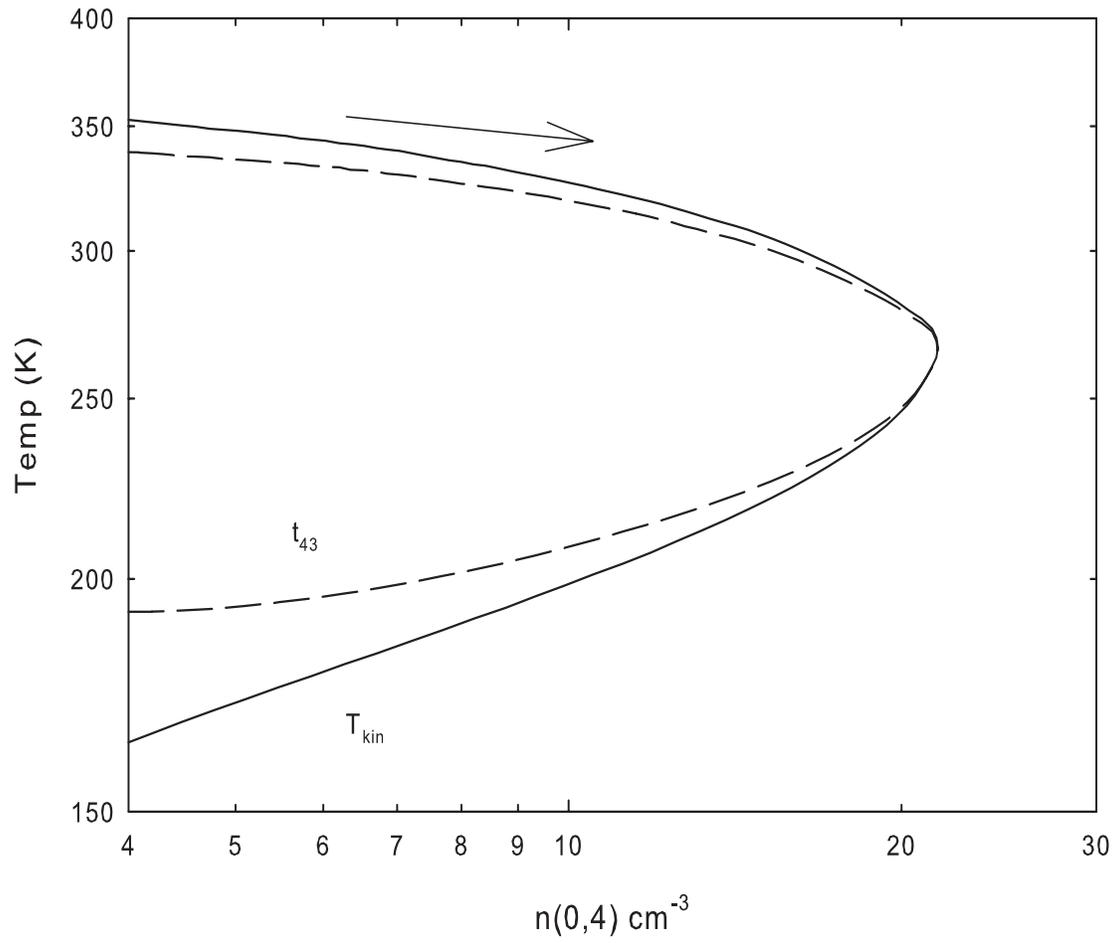

Figure 3. The excitation temperature $t_{43}$ (long-dashed line) and the kinetic temperature (solid line) as a function of $n(0,4)$. The arrow on the curves show the direction into the cloud.

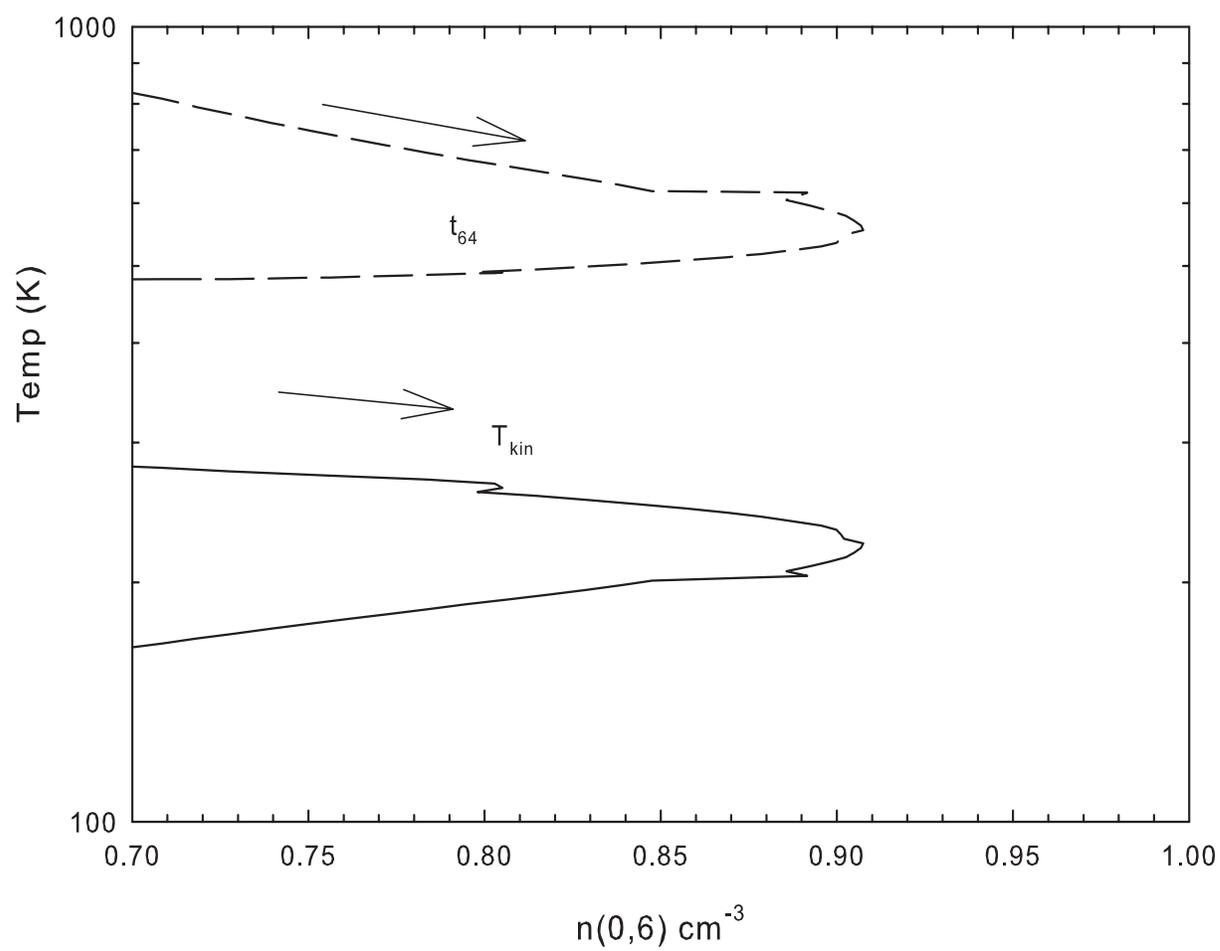

Figure 4. The excitation temperature $t_{64}$ (long-dashed line) and the kinetic temperature (solid line) as a function of $n(0,6)$. The arrows on the curves show the direction into the cloud.

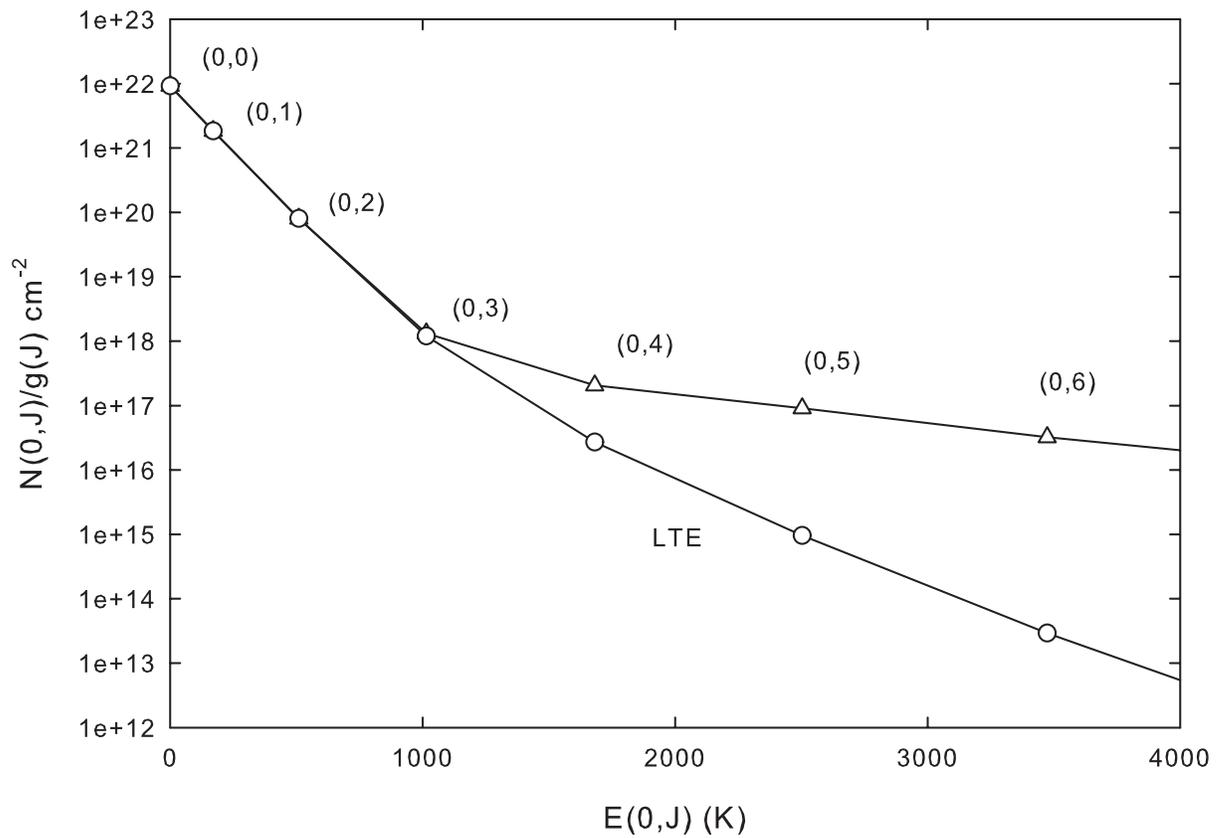

Figure 5a. The column densities of the $H_2$ rotational levels $J$ for v=0 as a function of excitation energy. LTE (circles), full $H_2$ model (triangles). The points marked LTE gives the column densities predicted for the derived temperature but with thermal level populations. If the gas were isothermal the points would fall along a straight line.

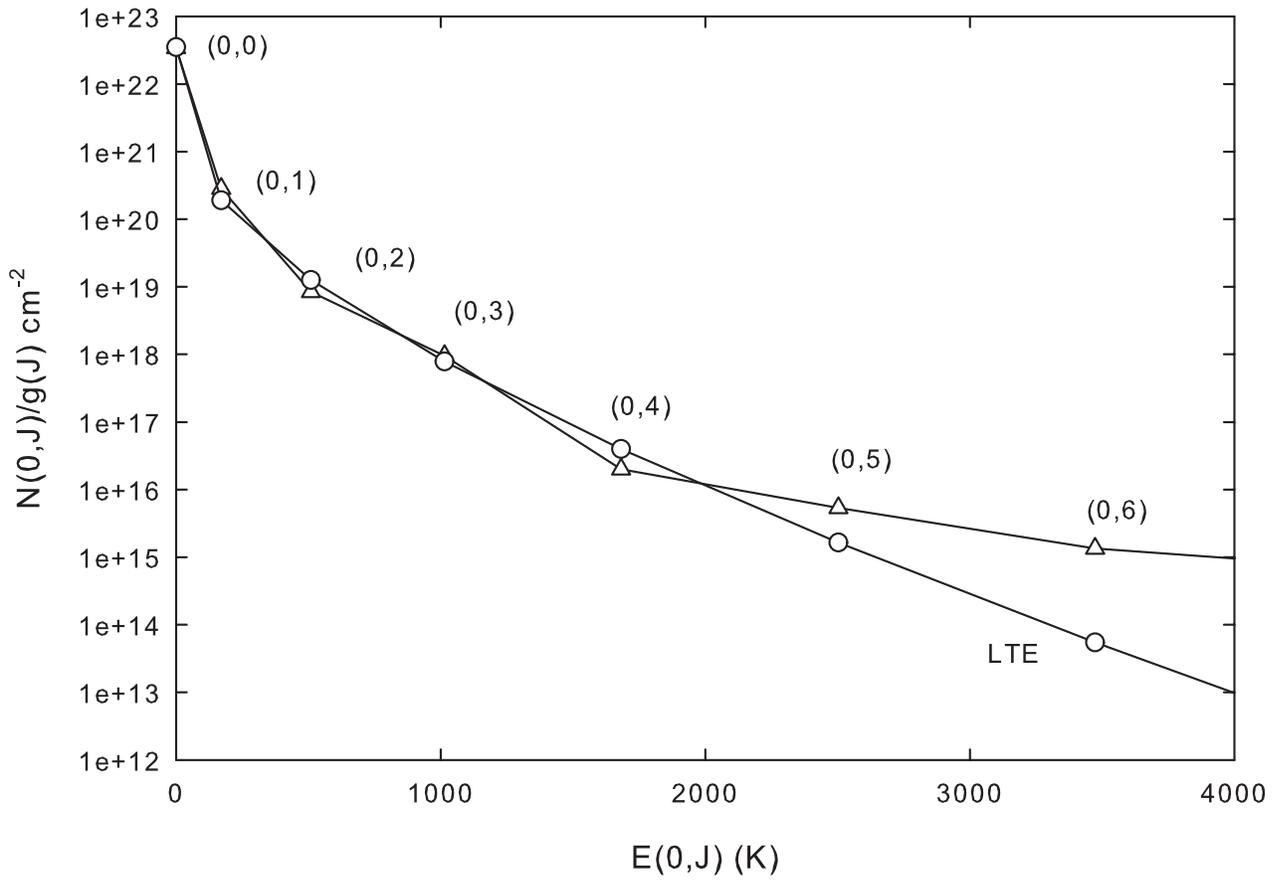

Figure 5b. Same as figure 5a, but without an enhanced cosmic-ray ionization rate.

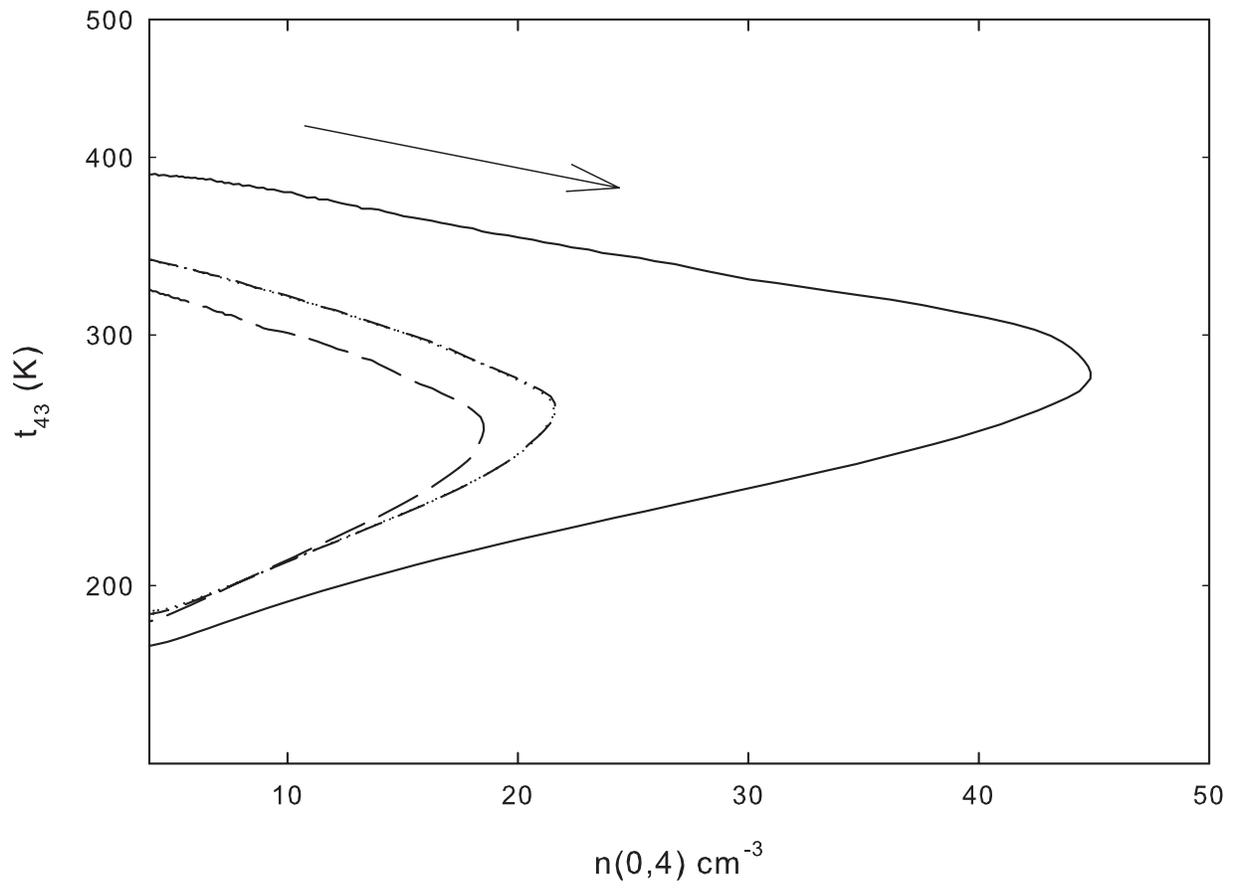

Figure 6. Same as Fig. 3, but using the $H_2$ collisional data sets given in Table 3.

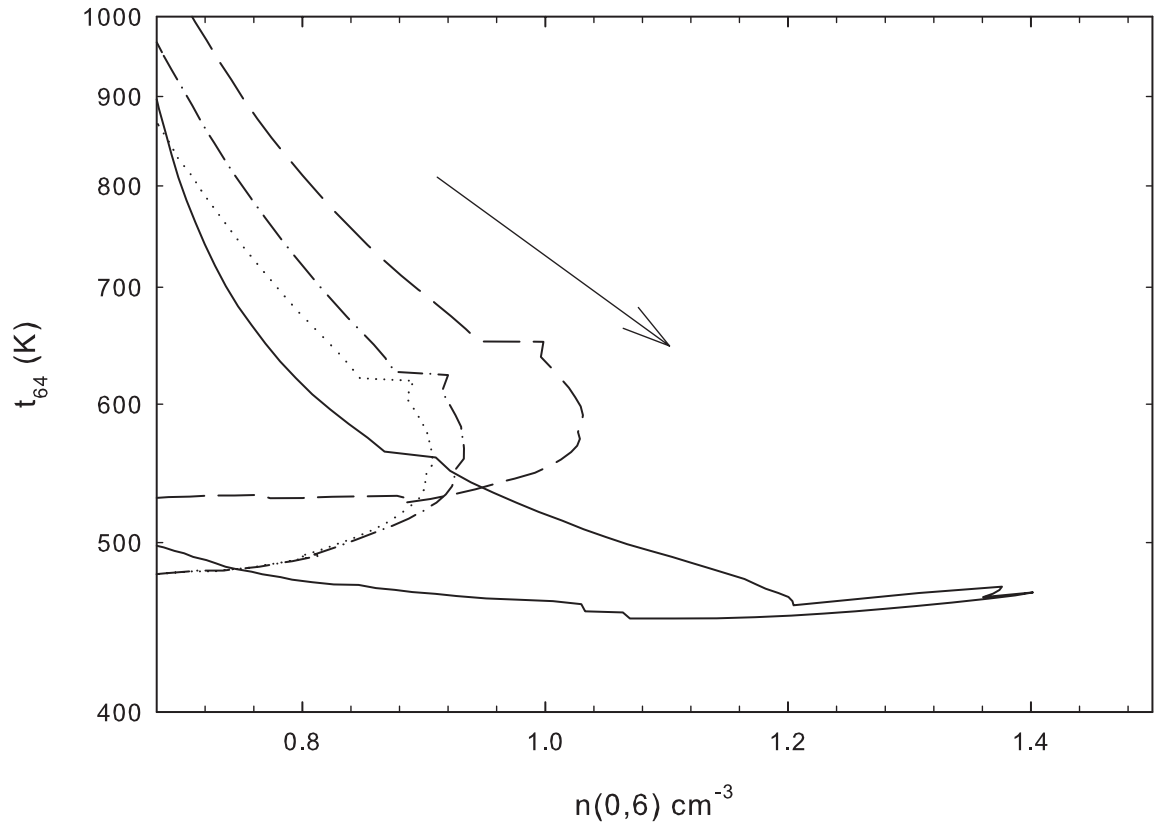

Figure 7. Same as Fig. 4, but using the $H_2$ collisional data sets given in Table 3.

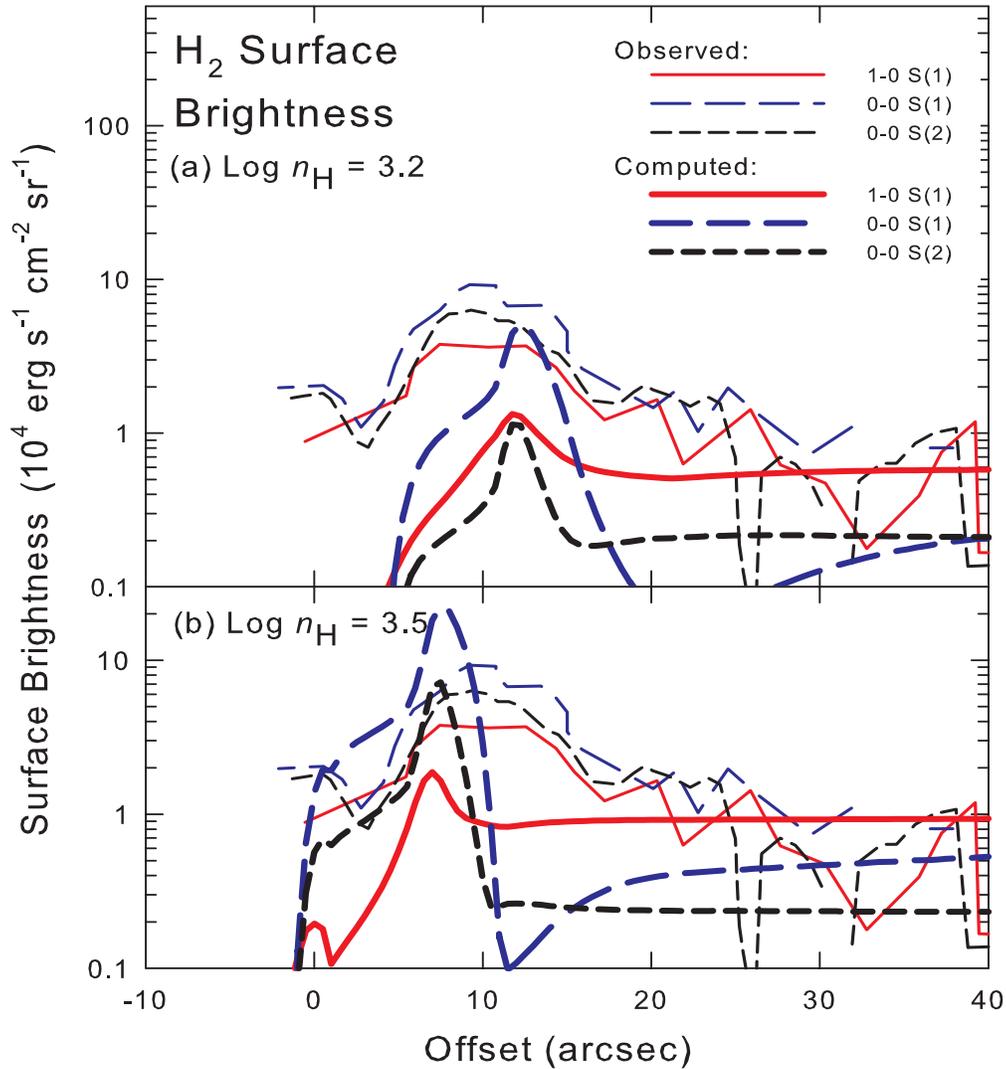

Figure 8a. Observed (thin lines) and computed (thick lines) H$_2$ surface brightness profiles for the 1-0 S(1), 0-0 S(1) and 0-0 S(2) lines. This shows the projected surface brightness computed by summing over a large number of individual rays passing through the tilted slab, as opposed to Fig. 2 which shows parameters along just a single ray. The offset scale is as observed on the sky, relative to the ionization front as delineated by the peak [S II] emission, with the starlight coming in from the left side of the figure. Note the reversal in the order of the peak surface brightness of the 1-0 S(1) and 0-0 S(2) lines as $n_H$, the gas density at the illuminated face, is increased.

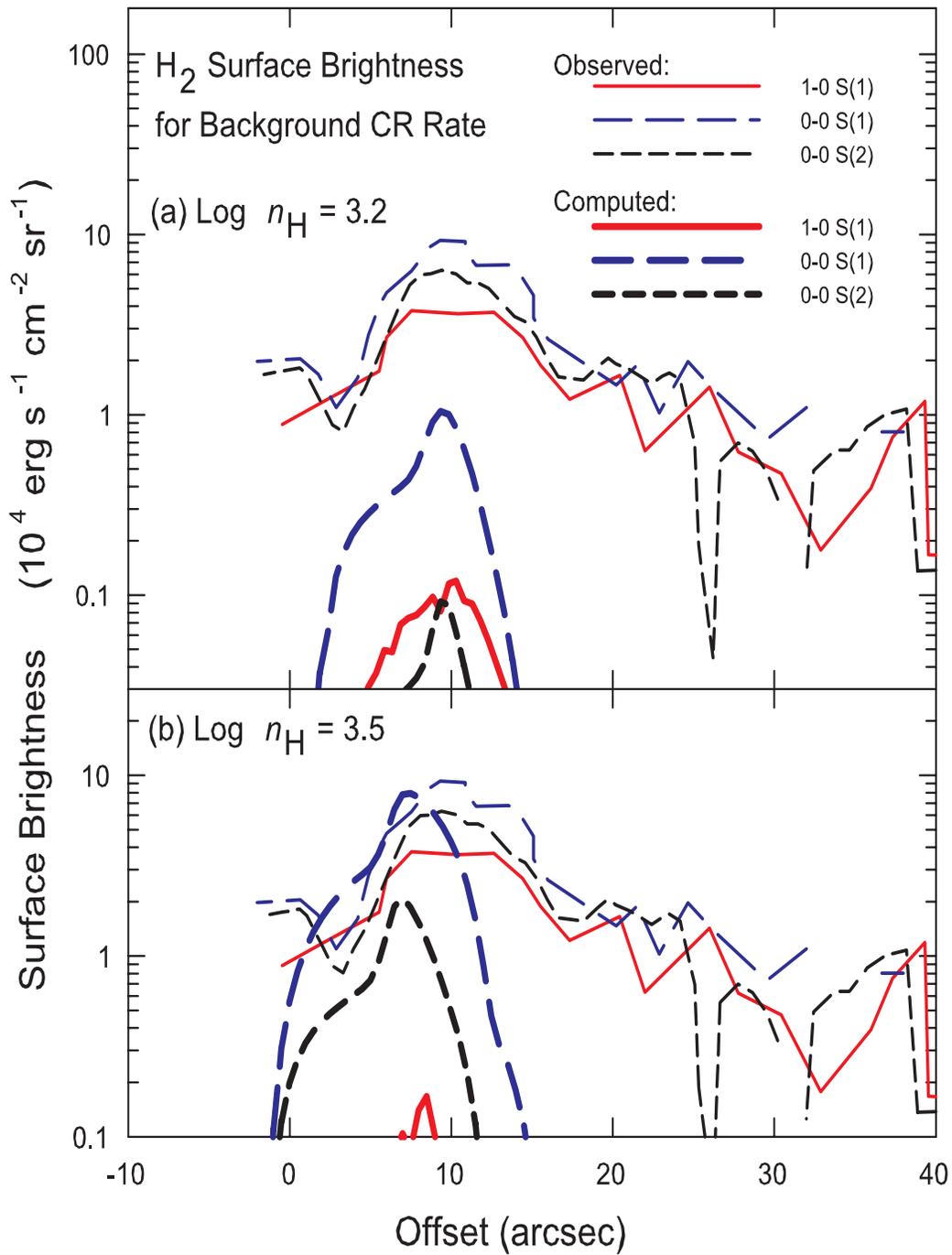

Figure 8b. Same as figure 8a, but without an enhanced cosmic-ray ionization rate.

# 5. Tables

**Table 1: Model parameters**

| Distance to the Orion nebula | 437±19 pc |
|---|---|
| Log $n_H$(illuminating face) | 3.2 cm$^{-3}$ |
| $R_v$ | 5.5 |
| PAH | $n_{PAH}/n_H^0 = 3\times10^{-7}$ |
| Magnetic Field | $<B> = 8$ µG at the illuminated face |
| Cosmic rays ionization rate of $H^0$ | Equipartition with magnetic field |
| Turbulence | 2 km s$^{-1}$ |
| Hard X rays | Bremsstrahlung emission at $10^6$ K, integrated luminosity $=10^{32.6}$ erg s$^{-1}$ |
| Stellar continuum of $\theta^1 C$ | Kurucz's stellar atmosphere calculations, $T_{eff}= 39700$K |
| CMB | $z = 0$ |

**Table 2: Gas phase chemical abundances by number**

| log He/H | -1.0223 |
|---|---|
| log C/H | -3.5229 |
| log O/H | -3.3979 |
| log N/H | -4.1549 |
| log Ne/H | -4.2218 |
| log Si/H | -5.3979 |
| log S/H | -5.0000 |
| log Fe/H | -5.5229 |
| log Mg/H | -5.5229 |
| log Cl/H | -7.0000 |
| log Ar/H | -5.5229 |

**Table 3: Model parameters**

| Case | Collider | | |
| --- | --- | --- | --- |
| | H | He | $H_2$ |
| S05 rates(solid line) | LeBo99 | LeBo99 | LeBo99 |
| S05 + new He rates (medium dashed line) | LeBo99 | Lee08b | LeBo99 |
| S05 + new He + new H rates (dash-dotted line) | Wrat07 | Lee08b | LeBo99 |
| Current rates (dotted line) | Wrat07 | Lee08b | Lee08a |

Notes: LeBo99 = Le Bourlot et al. (1999), Lee08a = Lee et al. (2008a), Lee08b = Lee et al. (2008b), Wrat07 = Wrathmall et al. (2007).

**Table 4. Rate coefficients at 500K for $H_2(0,8)$ to $H_2(0,6)$ transition.**

| | Colliders | | | |
| --- | --- | --- | --- | --- |
| | H | He | $H_2(0,0)$ | $H_2(0,1)$ |
| Le Bourlot et al. (1999) | $8.9\times10^{-14}$ | $1.9\times10^{-13}$ | $9.8\times10^{-14}$ | $1.3\times10^{-13}$ |
| Wrathmall et al. (2007). | $3.38\times10^{-13}$ | | | |
| Lee et al. (2008a) | | | $5.8\times10^{-14}$ | $7.4\times10^{-14}$ |
| Lee et al. (2008b) | | $1.6\times10^{-13}$ | | |